\documentclass[%
preprint,
superscriptaddress,
nofootinbib,
 amsmath,amssymb,
 aps,
]{revtex4-2}

\usepackage{graphicx}
\usepackage{xcolor} 
\usepackage{subfigure}

\usepackage{amsmath,amssymb}
\usepackage{amsfonts}
\usepackage{natbib}
\usepackage{amssymb}
\usepackage{booktabs}
\usepackage{mathtools}

\begin{document}

\title{Classifying basins of attraction using the basin entropy}

\author{Alvar Daza}
 \affiliation{Nonlinear Dynamics, Chaos and Complex Systems Group, Departamento de F\'{i}sica, Universidad Rey Juan Carlos, M\'{o}stoles, Madrid, Tulip\'{a}n s/n, 28933, Spain}
\email{alvar.daza@urjc.es}

\author{Alexandre Wagemakers}
 \affiliation{Nonlinear Dynamics, Chaos and Complex Systems Group, Departamento de F\'{i}sica, Universidad Rey Juan Carlos, M\'{o}stoles, Madrid, Tulip\'{a}n s/n, 28933, Spain}

\author{Miguel A.F. Sanju\'{a}n}
 \affiliation{Nonlinear Dynamics, Chaos and Complex Systems Group, Departamento de F\'{i}sica, Universidad Rey Juan Carlos, M\'{o}stoles, Madrid, Tulip\'{a}n s/n, 28933, Spain}

\date{\today}

\begin{abstract}
A basin of attraction represents the set of initial conditions leading to a specific asymptotic state of a given dynamical system. Here, we provide a classification of the most common basins found in nonlinear dynamics with the help of the basin entropy. We have also found interesting connections between the basin entropy and other measures used to characterize the unpredictability associated to the basins of attraction, such as the uncertainty exponent, the lacunarity or other different parameters related to the Wada property.
\end{abstract}

\maketitle

\section{Introduction}
Multistability is a common phenomenon in nature~\cite{pisarchik2014control}. From a deterministic point of view, it implies the existence of different asymptotic states in phase space for different choices of initial conditions. The set of all the initial conditions leading to one particular fate, such as an attractor or an exit of the system, is called a basin of attraction or an escape basin~\cite{nusse1996basins}. Since the early days of chaos theory, basins have been the object of extensive study.  For example, one of the many facets of dissipative chaos is the fractal nature of the basin boundaries~\cite{mcdonald1985fractal,aguirre_fractal_2009}. Along the last decades, a plethora of different kinds of basins with special characteristics has been reported: locally connected and disconnected~\cite{mcdonald1985fractal}, Wada~\cite{kennedy_basins_1991}, riddled~\cite{alexander_riddled_1992,ott_scaling_1993}, intermingled~\cite{lai1995intermingled}, etc. At the same time, a variety of different tools has arisen to describe and characterize them: the uncertainty exponent~\cite{grebogi_final_1983}, the lacunarity~\cite{mandelbrot1995measures}, the basin stability~\cite{menck_how_2013}, the Wada parameter~\cite{daza_testing_2015}, the Wada index~\cite{saunoriene2021wada}, etc.

Among them, the basin entropy~\cite{daza2016basin} integrates some previous measures in a natural way. It is extremely useful to characterize the unpredictability associated to the basins in different scientific contexts~\cite{nieto2020measuring, halekotte2020minimal, mugnaine2019basin, daza2017chaotic,zotos2018basins}. The present paper analyzes the different kinds of basins (Wada, riddled, intermingled, etc.) under the viewpoint of the basin entropy. From this inquiry, a classification emerges where iconic types of basins maximize one aspect of this measure. Needless to say, the basin taxonomy proposed here is just one of the many that can be chosen attending to different criteria. Nevertheless, it is striking to observe how well the paradigmatic basins of attraction fit into this classification. \textcolor{black}{Probably, the main contribution of the present work is to provide a framework to understand better the unpredictability associated to the different types of basins. Such framework is based on the basin entropy, and goes beyond of the mere quantification of the unpredictability: it lays the ground to a deeper understanding of concepts such as fractality and smoothness, Wada boundaries, riddled basins, symmetry of the basins and more.}

The paper is organized as follows. In Sec.~\ref{sec:basin_entropy}, we review the basics of the basin entropy, its definition and interpretation. Sections~\ref{sec:uncertainty_exponent}-\ref{sec:NA} are devoted to analyze the role of each ingredient of the basin entropy and its relation to some of the most common types of basins. We also dissect the relation of the basin entropy with other relevant measurements for each case. Finally, in Sec.~\ref{sec:discussion}, we summarize our results providing a rational classification of the basins according to their unpredictability, as measured by the basin entropy.

\section{Basics of the basin entropy}
\label{sec:basin_entropy}

Since its appearance~\cite{daza2018basin}, the basin entropy has been prolifically used in different contexts and with different aims. For instance, it is an efficient alternative to the box-counting algorithm for estimating the Hausdorff dimension~\cite{GUSSO2021111532}. It also has been successfully applied to the detection of KAM islands in conservative systems~\cite{nieto2020measuring}. Furthermore, it has been applied to investigate features of spatial patterns emerging in networks with competing species, using a model inspired in the rock-paper-scissors game~\cite{mugnaine2019basin}.

Among other purposes, the basin entropy has been a response to solve the vagueness of some affirmations concerning the unpredictability associated to either basins of attraction or exit basins. For example, Wada basins were typically assumed to be {\it more unpredictable} than fractal basins without the Wada property~\cite{aguirre_wada_2001}. Certainly, such kind of statements could have been received a more rigorous consideration. Precisely, these considerations lead to a central question: how can we quantify the intuitive notion of the unpredictability associated to the basins of attraction? The basin entropy provides some answers.

Here, we will sketch the definition of the basin entropy and some of the main results explained in~\cite{daza2018basin}. The basins of attraction can be considered as a coloring of the phase space where each initial condition is matched to a color corresponding to an attractor. For the definition of the basin entropy, we use a covering of size $\varepsilon$ of the basins. The proportion of colors in each $\varepsilon$-ball defines the probability associated to each color $p_i$. Then, the Gibbs entropy of each ball can be computed as $S_i=-\sum_{i=1}^{N_A} p_i\log{p_i}$, where $N_A$ is the number of different colors in the basin. The basin entropy is defined as the average of the Gibbs entropy for all balls $S_b= \langle S_i \rangle$.

Although computing the basin entropy requires no extra assumptions but just applying the instructions of the previous paragraph, we can delve deeper into its meaning with the following analysis. Using the definition of the box-counting dimension, the number of $\varepsilon$-balls covering a phase space of dimension $D$ grows as $N=n \varepsilon ^ D$. This number, for the covering of a basin boundary of dimension $D_k$, grows as $N_k=n_k \varepsilon ^ {D_k}$. Also, we can assume that the basins are equally distributed around their boundaries with identical probabilities $p_k$. The entropy of an $\varepsilon$-ball in such a boundary is $S_i = - \log ~ m_k$, where $m_k\geq 2$ refers to the number of basins separated by the boundary. The sum of all $S_i$ for the boundary $k$ is equal to $N_k \log~ m_k$. Bringing everything back together, we obtain the basin entropy
\begin{align}
    S_b=\sum\limits_{k=1}^{k_{max}}  \dfrac{n_k}{ n} \varepsilon^{\alpha_k} \log (m_k),
    \label{eq:Sb3}
\end{align}
where $\alpha_k=D-D_k$ is the uncertainty exponent of the $k$ boundary. As explained in~\cite{daza2018basin}, this expression reveals the main ingredients that contribute to the basin entropy and therefore to the uncertainty associated to the long-term prediction of the associated system. In the following, we show the role of each of these ingredients in our attempt to classify the most common types of basins of attraction found in nonlinear dynamics.

\section{Uncertainty exponent: smoothness of the boundaries}
\label{sec:uncertainty_exponent}
The first obvious classification of basin boundaries is between smooth and fractal. This distinction can be analyzed in terms of the uncertainty on a $\varepsilon$ neighborhood of an initial condition. If all the initial conditions in the neighborhood lead to the same attractor, this initial condition is called certain. If at least two initial conditions lead to distinct attractors, this is an uncertain initial condition. For an average over the phase space, the ratio $f$ of uncertain to the certain initial conditions is a function of $\varepsilon$. The usual interpretation of $f(\varepsilon)$ is a measure of the ignorance on the final state of initial conditions in the boundary. Smooth basins are well behaved in the sense that, if the initial uncertainty $\varepsilon$ is decreased, $f$ decreases proportionally. This is probably the expected behavior, but this is not what happens with fractal boundaries. In this case, the ratio $f$ decreases as $f\sim \varepsilon^{D-d} = \varepsilon^{\alpha}$, being $D$ the topological dimension of the phase space and $d$ the box-counting dimension of the basin boundary. As a consequence, fractal boundaries have $0\leq\alpha<1$, so one might decrease the uncertainty of the initial conditions $\varepsilon$ by a half (e.g. doubling the resolution of the grid of points), and still get a ratio of uncertain initial conditions $f$ very similar to the original one. Indeed, in the extreme cases when $\alpha=0$, $f$ remains constant for all the values of $\varepsilon$. This is exactly what happens for the case of riddled basins~\cite{alexander_riddled_1992}.

Therefore, the uncertainty exponent $\alpha$ measures the rate of growth of the space occupied by the boundary at different magnification scales. This parameter $\alpha$ does not only allows us to discern between smooth and fractal boundaries, but also conveys information about how the boundary fills the phase space. The closer $\alpha$ is to zero, the closer the boundary is to fill completely the phase space. However, as we shall see later, the uncertainty exponent fails to grasp other relevant aspects of the basin boundaries.

The uncertainty exponent of each boundary appears explicitly in Eq. \ref{eq:Sb3}, so the basin entropy also captures the scaling of the boundaries. However, it can also provide an efficient test of fractality when the basin is computed with at a single scale. The solution comes as a test \cite{puy2021test} able to distinguish between smooth boundaries and fractal basins. This test, based on the basin entropy, cannot give information about the rate of growth of the boundary, but can reveal the existence of structures visible at a given resolution. The principle of detection is to compute the basin entropy only at the boundary as accurately as possible with a ball of radius $\varepsilon$. If the value lies within a theoretical interval, the boundary is smooth. Otherwise, it means that it has structures below the scale $\varepsilon$.

\section{Lacunarity: connectedness of the basins}

The factor $n_k/n$ of Eq.~\ref{eq:Sb3} can be identified with the size of the boundaries, which is related to the concept of lacunarity~\cite{mandelbrot1982fractal}, although it is not quite the same. To illustrate the meaning of the lacunarity and its role in the basin entropy we can think of a simple example. Assume we have two different systems, both with two attractors $N_A=2$ and both with smooth basin boundaries, so that $D_k=1$. The basin entropy can still be different if, for example, the basin of an attractor is just a black disk or if it consists of $N$ disks with the same total area. As long as we use disks or other smooth shapes, we would have the same uncertainty exponent $\alpha=1$. Furthermore, if we carefully keep the total area of all the disks unaltered, the basin stability~\cite{menck_how_2013} would remain constant as well. Having just one disk with area $\pi R^2$ means having a boundary of length $l=2\pi R$, while having $N$ disks with total area $\pi R^2$ implies a perimeter $l_i=2\pi R/\sqrt{N}$ for each disk. The total length of the boundary in this latter case can grow indefinitely as we keep increasing the number of circles $l_T=\sum_{i=1}^{N} l_i= 2\pi R \sqrt{N}$. This difference in the uncertainty has its origin in their different lacunarity.

\begin{figure}[!h]
    \centering
    \subfigure[]{\includegraphics[width=0.45\linewidth]{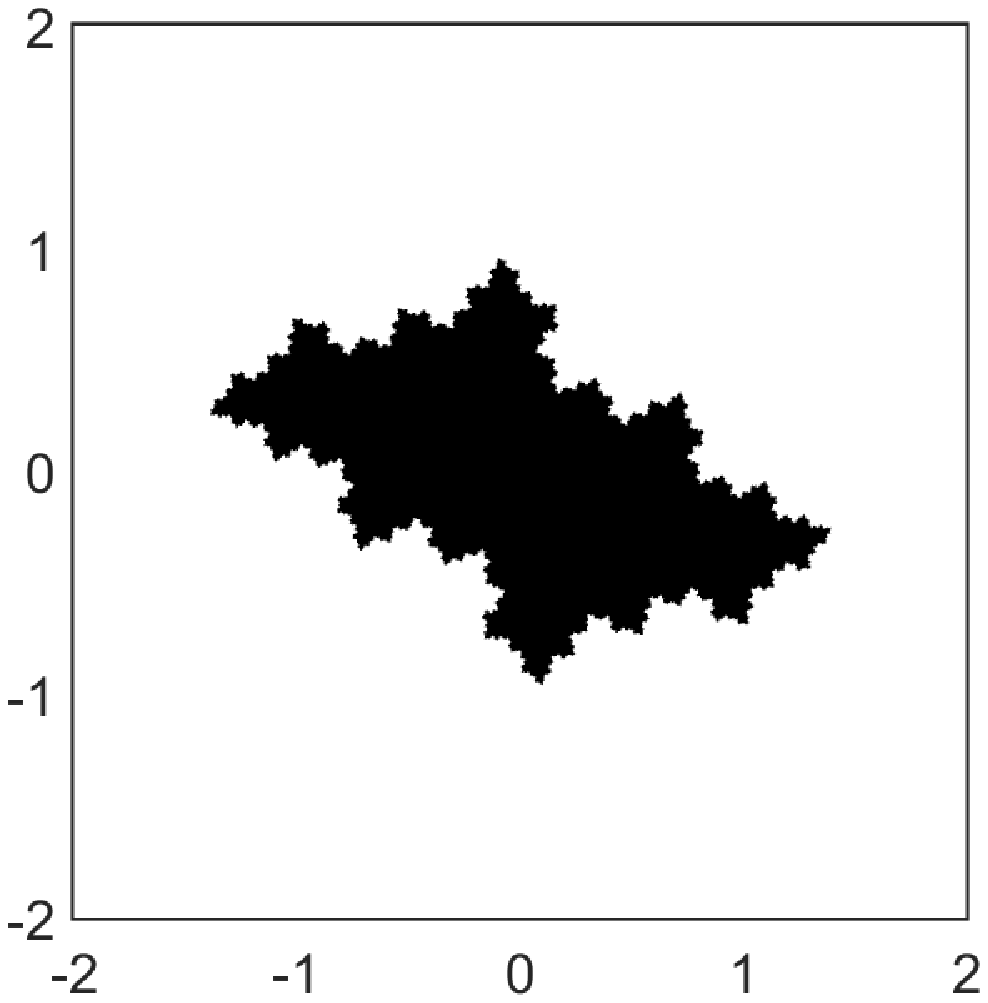}}
    \subfigure[]{\includegraphics[width=0.45\linewidth]{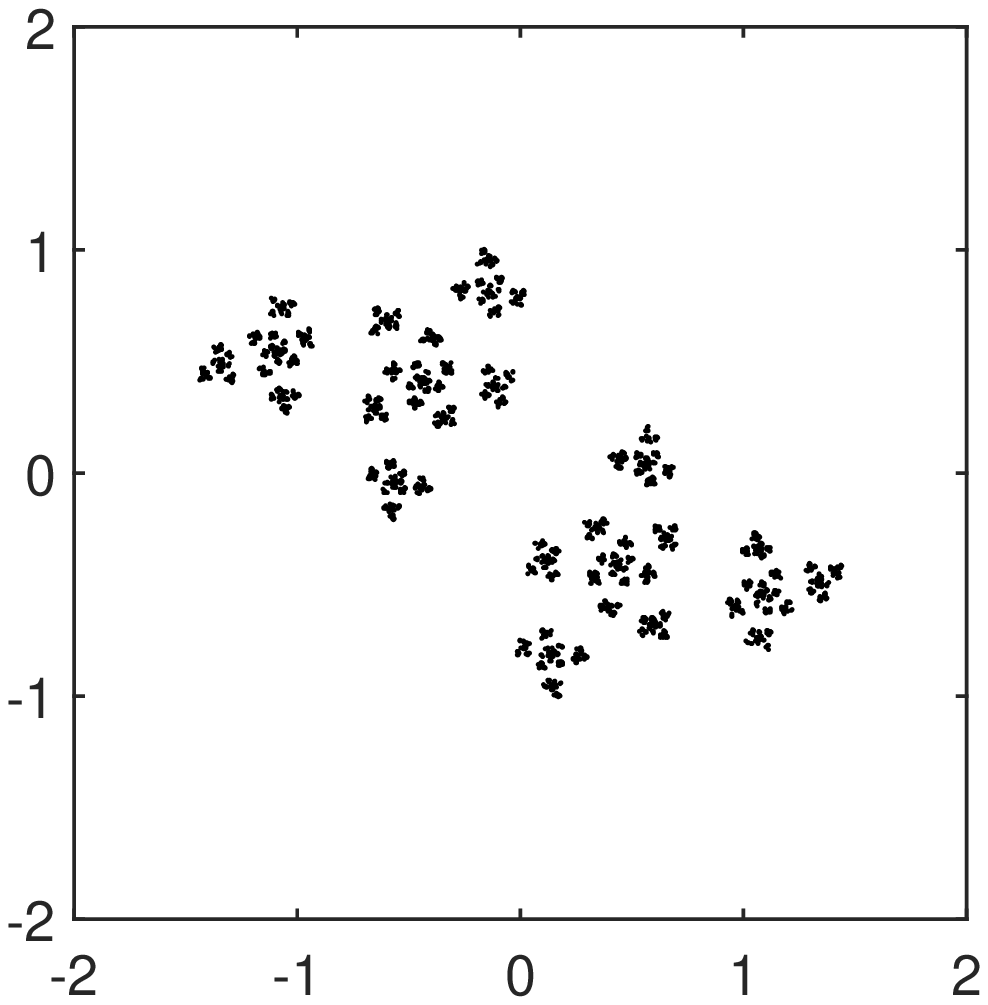}}
    \caption{Basins of attraction of the map $z_{n+1}=z_n^2+c$ for parameters $c=-0.4849 + 0.4472i$ \textcolor{black}{(left)} and $-0.4548 + 0.7688i$ \textcolor{black}{(right)}. The two figures have fractal boundaries with the same uncertainty exponent $\alpha=0.8$.}
    \label{fig:Julia1}
\end{figure}

\begin{figure}[!h]
    \centering
    \subfigure[]{\includegraphics[width=0.4\linewidth]{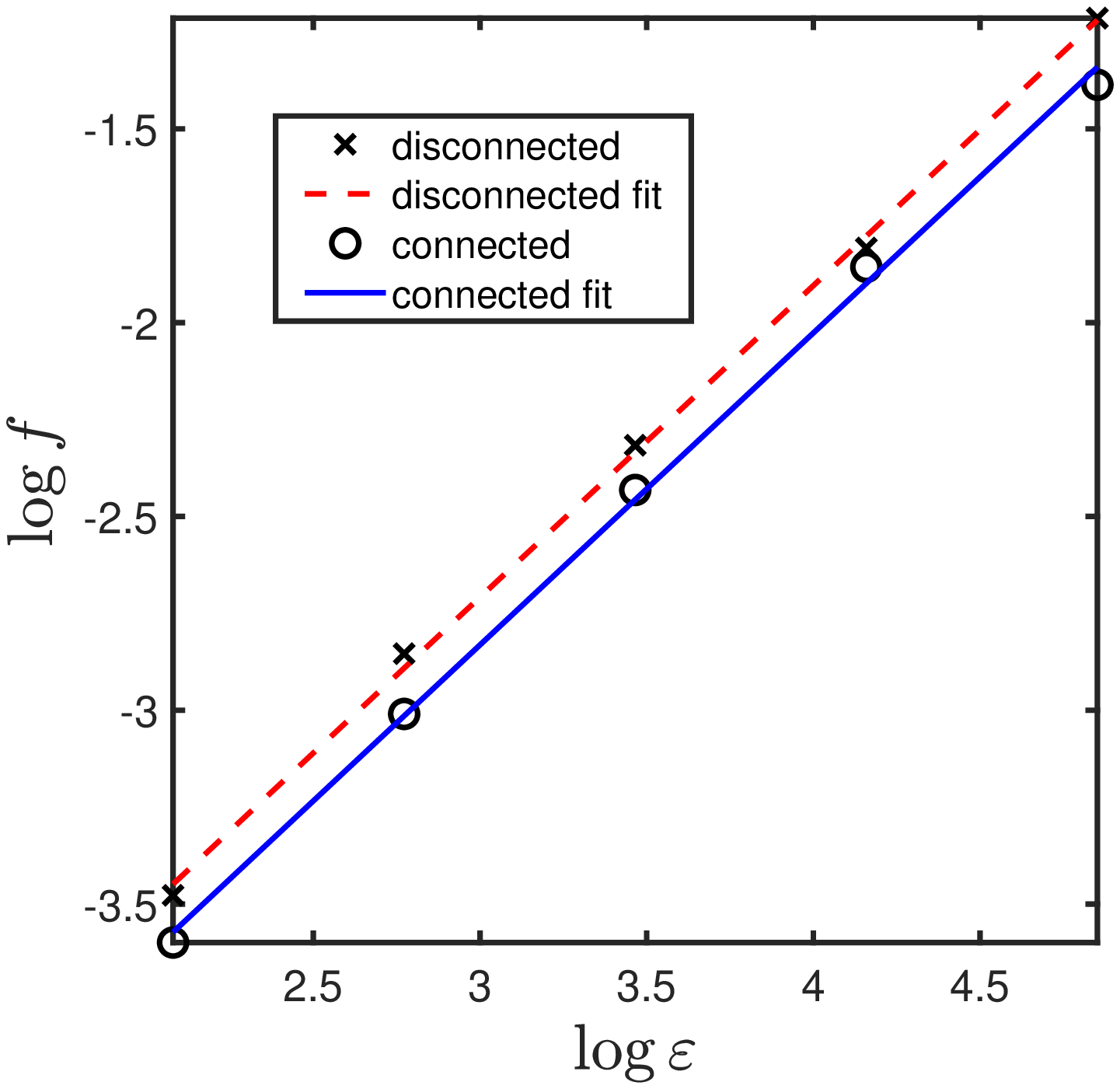}}
    \subfigure[]{\includegraphics[width=0.45\linewidth]{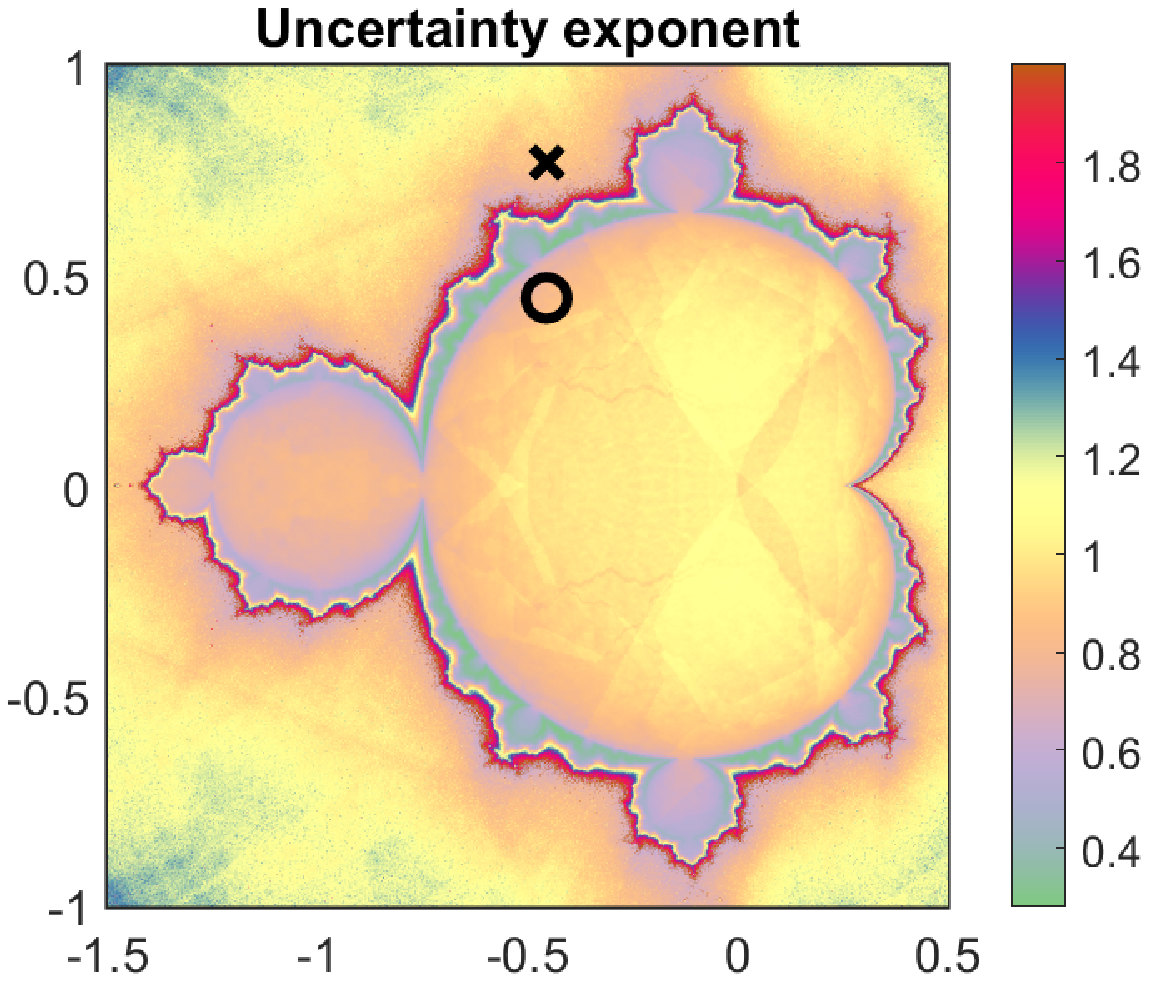}}
    \subfigure[]{\includegraphics[width=0.45\linewidth]{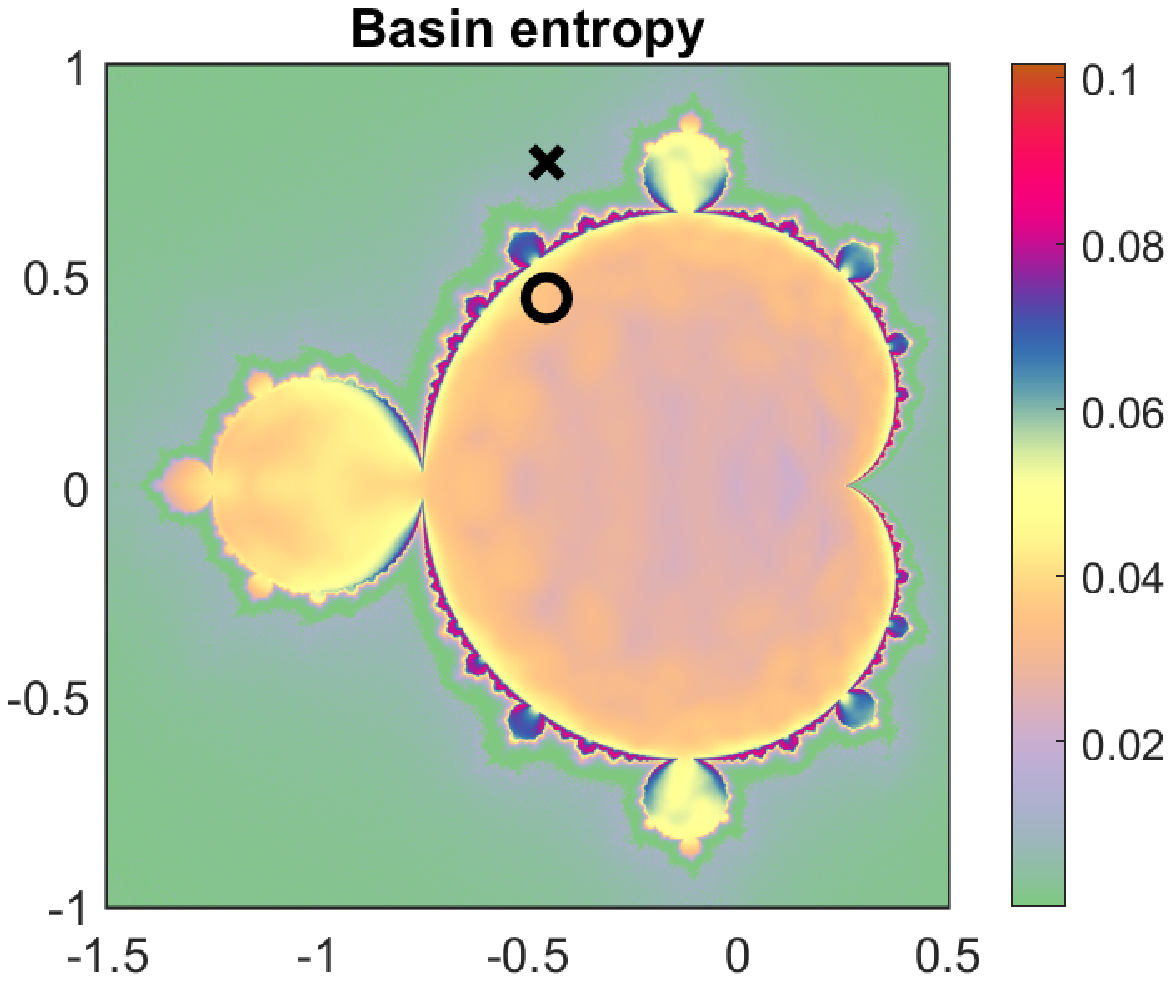}}

    \caption{
    (a) Uncertainty dimension calculation for both Fatou sets in Fig.~\ref{fig:Julia1}. Both of them have the same uncertainty exponent (parallel curves) but their lacunarity is different. 
    (b) Heat map of the uncertainty exponent for different values of the complex parameter $c$ for the Fatou sets. \textcolor{black}{Although the uncertainty exponent is typically between 0 and 1, here it can take larger values because we are dealing with Cantor-dust structures, as explained in the text.} (c) Same as (b), but here the color represents the basin entropy. While for (b) many points inside and outside the Mandelbrot set have the same color (same uncertainty exponent for connected and disconnected sets, alike), the basin entropy is clearly different for points that belong to the Mandelbrot set. \textcolor{black}{In panels (b) and (c), the values corresponding to the Fatou sets of Fig.~\ref{fig:Julia1}-(a) and (b) are represented as a circle $(c=-0.4849 + 0.4472i)$  and a cross ($c=-0.4548 + 0.7688i$)}.}
    \label{fig:Mandelbrot}
\end{figure}

For fractal boundaries, the lacunarity can be very different from one basin to another. To investigate its role more in detail we will focus on basins with the same uncertainty exponent. For example, the Fatou sets in Fig.~\ref{fig:Julia1}(a) and Fig.~\ref{fig:Julia1}(b) have this property, but one is a connected set and the other one is disconnected. In Fig.~\ref{fig:Mandelbrot}(c), we use a colormap to show the uncertainty exponent of Fatou sets with different values of the complex parameter $c$, which produces a picture that resembles the Mandelbrot set, as expected. We can observe that the same colors are inside and outside the Mandelbrot set, meaning that the same values of the uncertainty exponent $\alpha$ can be found for connected and disconnected sets, alike. Thus, the uncertainty exponent provides no information about the connectivity of the basins, as stated for example in~\cite{falconer2004fractal}.

The Fatou sets depicted in Fig.~\ref{fig:Julia1} have the same uncertainty exponent but different lacunarity. The sparse structure of the disconnected set indicates a higher lacunarity than the connected set. Figure~\ref{fig:Mandelbrot}(a) represents $f$ versus $\varepsilon$ for the two basins in a log-log scale. Since the slope of the linear fit estimates the uncertainty exponent, the parallel plots reveals the same fractal dimension for the two examples. The lacunarity is sometimes interpreted as the intercept of the linear fit on the vertical axis. This explains why the plot corresponding to the disconnected Fatou with higher lacunarity (in red) is above the plot of the connected set (in blue). The basin entropy is larger for the connected basin than for the disconnected basin for all scales.

Although the lacunarity of the disconnected set is higher, the probability of landing in one of its islands is very small. Indeed, from a strictly mathematical perspective, the basin entropy of all the disconnected Fatou sets should be equal to zero, because they are zero Lebesgue measure sets (two-dimensional Cantor dusts) and the probability of randomly landing exactly on a point belonging to them is exactly zero. However, if we represent these disconnected Fatou sets with finite resolution, such as in Fig.~\ref{fig:Julia1}(b), their basin entropy is different from zero and it conveys some information about their structure. For this same reason, the values of the basin entropy for the connected Fatou sets are much larger than for the disconnected ones, as can be clearly seen in Fig.~\ref{fig:Mandelbrot}(c).

\begin{figure}[!h]
    \centering
    \subfigure[]{\includegraphics[width=0.45\linewidth]{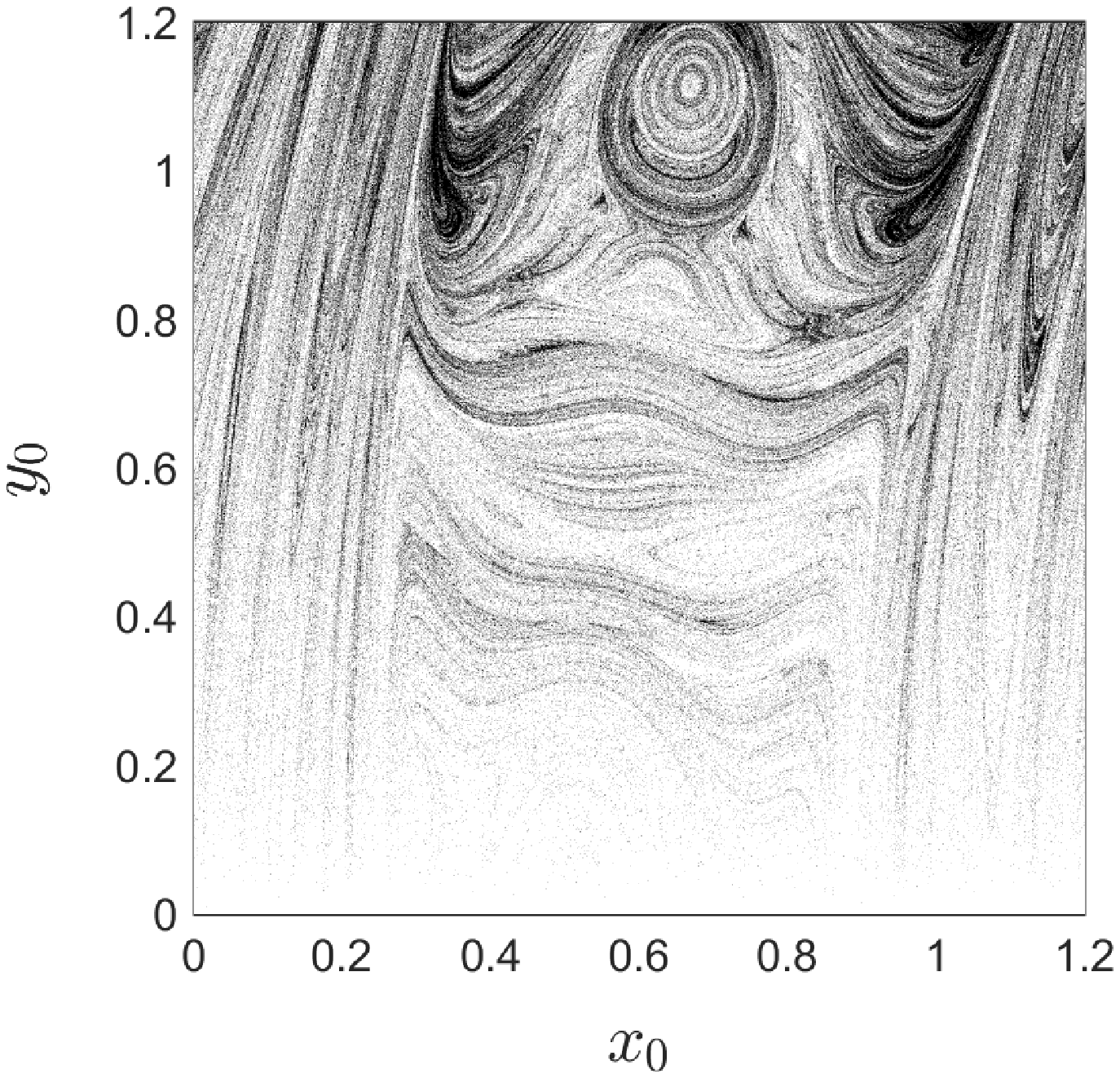}}
    \subfigure[]{\includegraphics[width=0.45\linewidth]{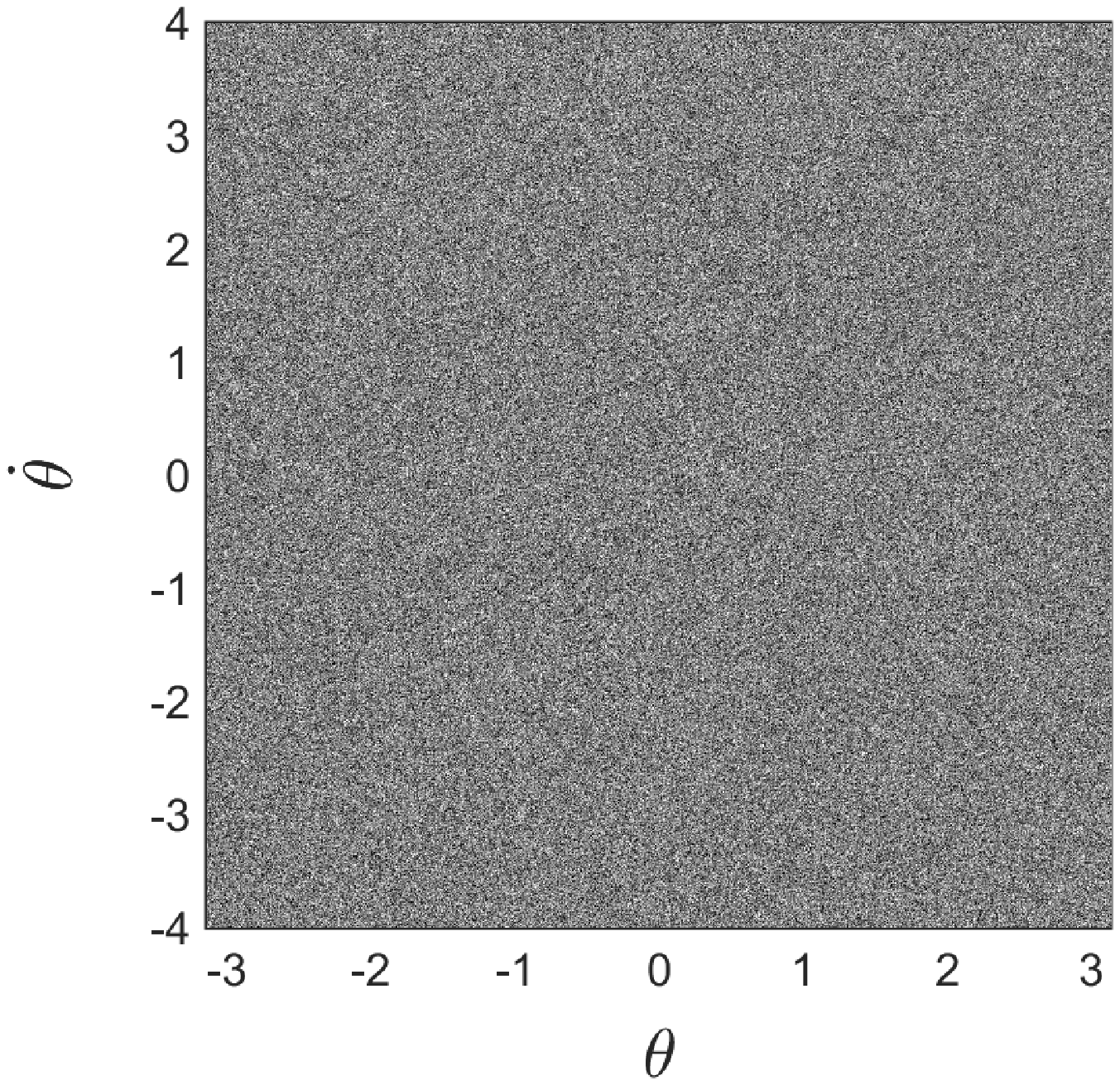}}
    \caption{ In (a), we reproduce the basins published in \cite{ott1994transition} that have the riddled property. In (b) we display the basins of the forced damped pendulum $\ddot x + 0.2 \dot x + \sin(x) = 1.3636 \sin 0.5 t$, which is also riddled. While the two basins have the same uncertainty exponent, their fine structure clearly differs as can be observed from the figures.
    \label{fig_riddle}}
\end{figure}

In the previous example, both basins had the same uncertainty exponent although there was a topological distinction between connected and disconnected boundaries. In Fig.~\ref{fig_riddle}, the two basins from different dynamical systems have an uncertainty exponent $\alpha = 0$~\footnote{In this article, $\alpha = 0$ is considered a sufficient condition to decide if a basin is riddled, see~\cite{alexander_riddled_1992, ott_scaling_1993, ott1994transition, aguirre_fractal_2009} for further discussions on this topic.}. For both boundaries, the disconnected set fills all the phase space. Nonetheless, there are striking differences in the texture of both basins.

The basin entropy is an average of the Gibbs entropy for boxes all over the phase space. The restriction of this average to boxes on the boundary is called the boundary basin entropy $S_{bb}$. For riddled basins, the boundary fills the phase space and both averages are the same: $S_b=S_{bb}$. Still, the distribution of basins inside each box may vary depending on the location of the box. Here we have to interpret the lacunarity in a wider sense that also includes the distribution of the basins inside each box. Figure \ref{fig_riddle}(a) was first described in Ref. \cite{ott1994transition} where the authors have proved the riddled nature of the basins. The computation of the basins of the forced damped pendulum in Fig. \ref{fig_riddle}(b) reveals riddled basins with an aspect very similar to white noise. The basin entropy obtained for Fig.~\ref{fig_riddle}(a) is $S_b \simeq 0.4$ and $S_b \simeq \log 2$ for Fig.~\ref{fig_riddle}(b). This difference in the basin entropy between the two figures reflects the fine structure of the basins that cannot be grasped with the fractal dimension. The uncertainty is lower in one case due to the asymmetry in the distribution of the basins of each attractor. Also, the uncertainty is maximum in the case of the forced pendulum since the value of the basin entropy is maximum (for two attractors).


\section{Number of attractors and Wada property}
\label{sec:NA}
The previous sections account for situations with two attractors, but in multistable systems a large number of attractors with their corresponding basins can coexist~\cite{feudel1996map}. Furthermore, fractal geometry allows for three or more basins to share a common boundary. Such situation is often referred to as Wada basins, and many works have been devoted to identify, prove and characterize the Wada property~\cite{daza_testing_2015,daza2018ascertaining,wagemakers2020detect,wagemakers2020saddle}. Its interest resides precisely in the particular uncertainty of these basins: any deviation of an initial condition lying in a Wada boundary can evolve into any of the three or more different fates of the system. This unique feature provokes an increase of the basin entropy since clearly the uncertainty grows when the number of different attractors increases.

\begin{figure}[!h]
    \centering
    \subfigure[]{\includegraphics[width=0.35\linewidth]{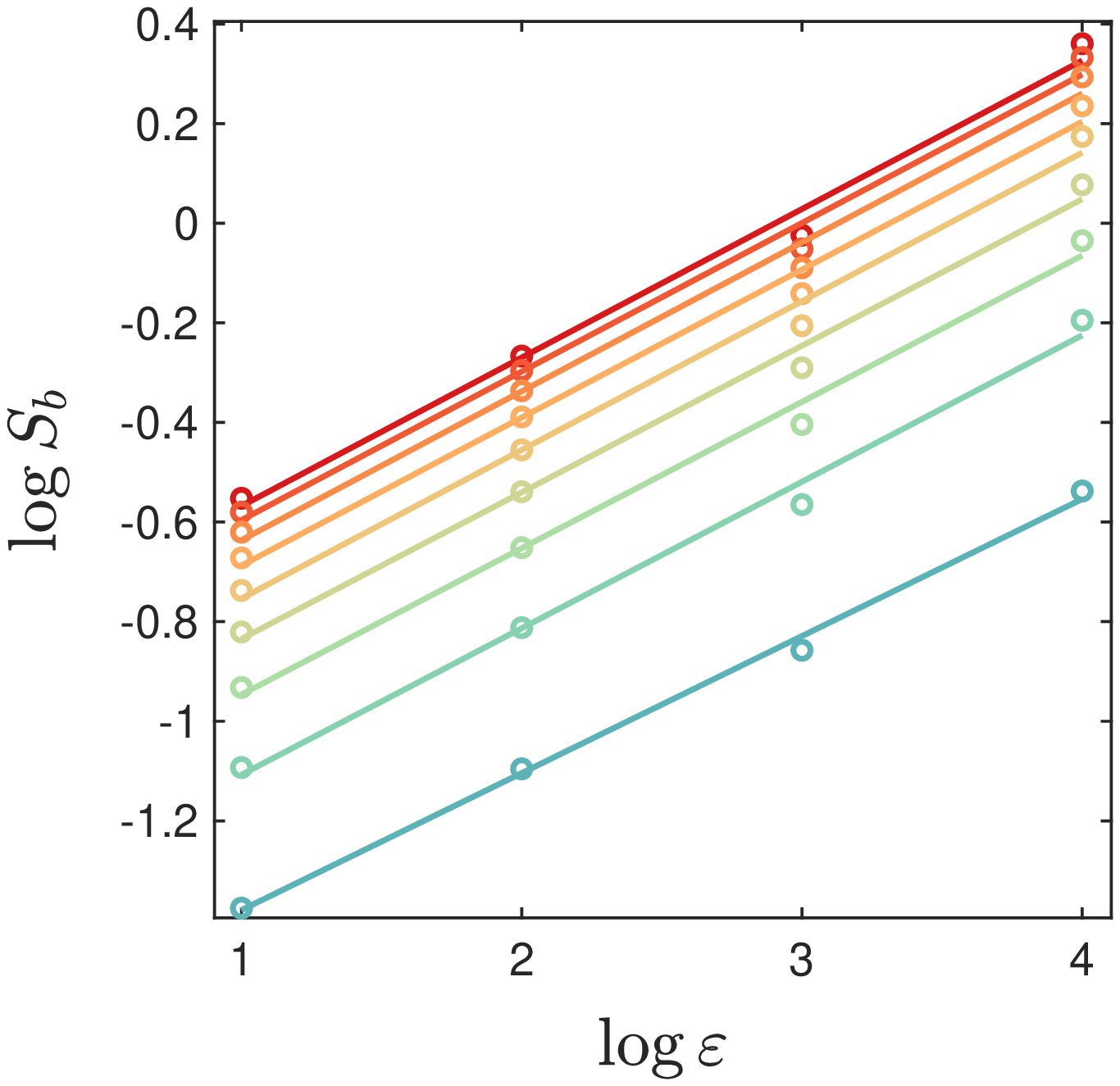}}
    \subfigure[]{\includegraphics[width=0.35\linewidth]{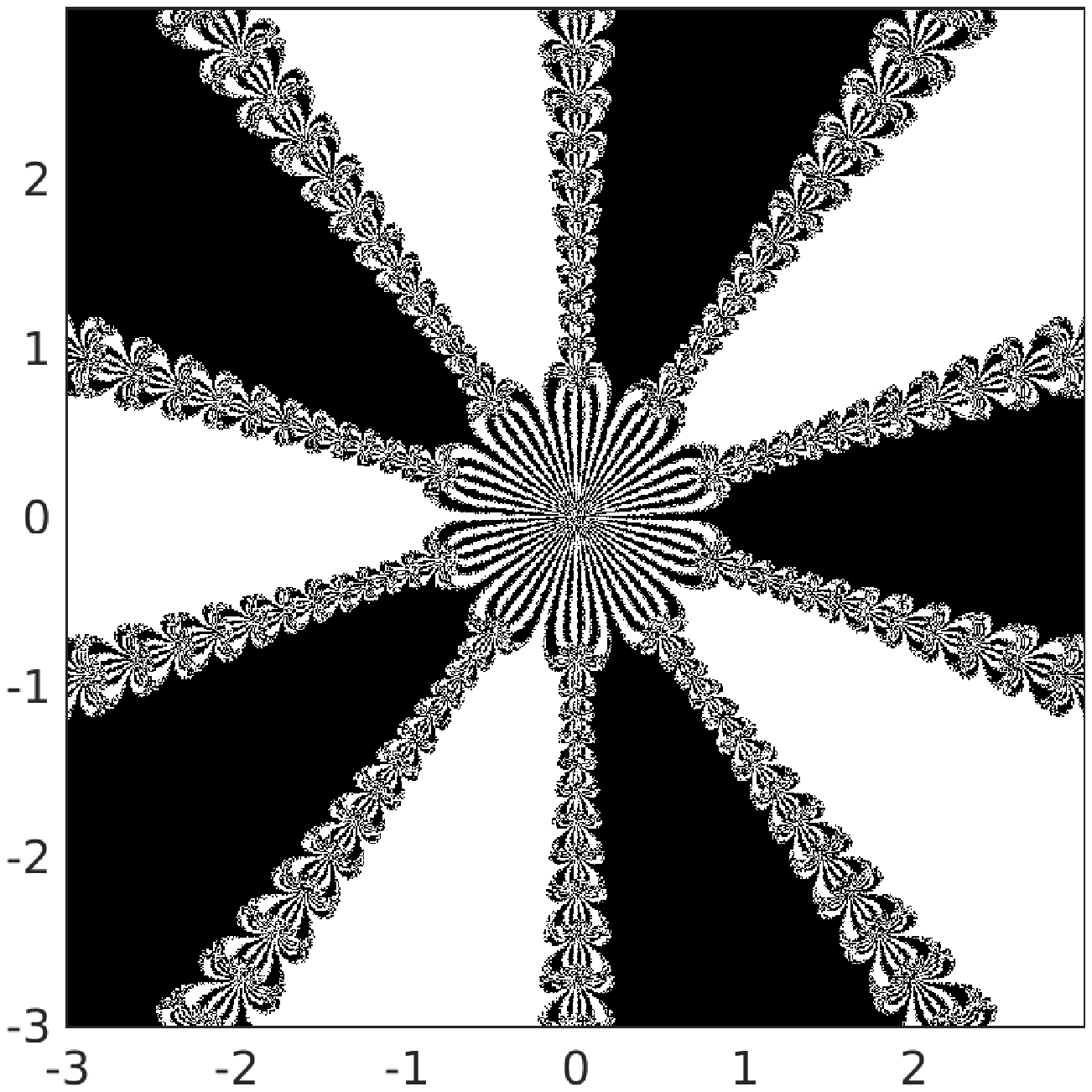}}
    \subfigure[]{\includegraphics[width=0.35\linewidth]{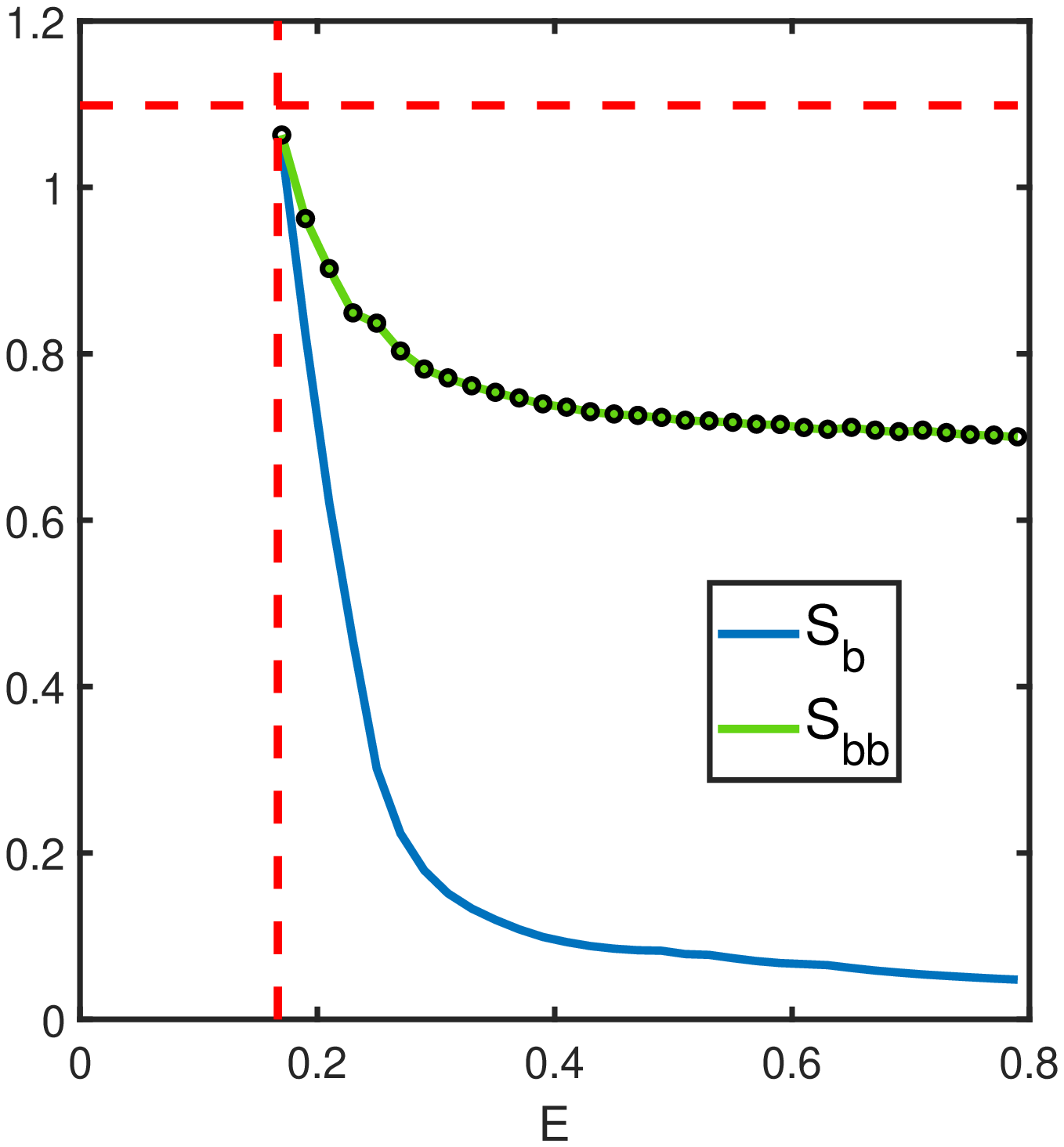}}
    \caption{
    (a) Basin entropy calculation for the Newton fractal with $10$ attractors. Following the merging method, the number of attractors $N_A$ varies from $2$ (bottom) to $10$ (top). The slope is always the same because the boundaries remain the same after the merging operation, but the slope increases with the number of attractors converging geometrically. (b) Newton fractal for $10$ attractors, where we have merged the basins creating a binary picture ($N_A=2$). (c) $S_b$ (in blue) and $S_{bb}$ (in green) calculation for different values of the energy in the H\'enon-Heiles system. As the critical energy $E=1/6$ is approached (vertical line), both values converge to $S_b=S_{bb}=\log 3$. We have also plotted the Wada index multiplied by $\log N_A/N_A$ as black circles, showing that for Wada boundaries Eq.~\ref{eq:wSbb} is fulfilled.}
    \label{fig:NA}
\end{figure}

The merging method~\cite{daza2018ascertaining}, one of the algorithms used to detect Wada basins, provides some insights into the relation of the number of attractors and the basin entropy. In Fig.~\ref{fig:NA}(a), the points on the top and its corresponding fitting line (both in red), show the basin entropy for the Newton fractal with ten roots, i.e., ten different basins. Following the procedure described in~\cite{daza2018ascertaining}, we have merged the basins one by one, so initially we had ten basins and at the end we only had two basins. A possible result of this merging process is shown in the basins of Fig.~\ref{fig:NA}(b). Interestingly, Wada boundaries remain unaltered by this merging procedure, so the only difference between the merged versions of the basins is the number of attractors $N_A$ (the number of different colors of the picture). This is clearly seen in  Fig.~\ref{fig:NA}(a), where in the log-log plot of the basin entropy $S_b$ versus the box size $\varepsilon$, the lines for the different merged basins have the same slope and they only differ by their intercept. Also, going from the bottom line of Fig.~\ref{fig:NA}(a) (corresponding to two attractors) to the top line (corresponding to ten attractors), we can see how the values of the basin entropy tend to a limiting value as the number of attractors increases. \textcolor{black}{Given that the basins of the Newton fractal have a single boundary $k=1$, we have that Eq.~\ref{eq:Sb3} simply becomes $S_b = (n_b/n) \varepsilon^\alpha \log m$.} This result is in agreement with Eq.~5 of \cite{saunoriene2021wada}, so that the unpredictability of the system increases more when it changes from $2$ to $3$ attractors than when it changes from $9$ to $10$.

Another important aspect of these calculations is related to symmetry. In this case, the symmetry of the basins of the Newton fractal implies the invariance of the basin entropy for any merging operation. However, this is not true in general. Even though the boundaries remain unaltered by the merging process, the basins do change. Since the basin entropy measures the unpredictability associated to the basins, if they are different, the value of the basin entropy will be different too.

Thus, the number of basins separated by the boundaries plays an important role in the basin entropy, but so does another important parameter: the Wada index~\cite{saunoriene2021wada}. This number accounts for the distribution of the colors inside the boundary boxes, so that it is maximal for equiprobable situations. Briefly, the Wada index $w$ for a given box of size $s$ in pixels can be computed as:
\begin{equation}
   w = \left\{ \begin{array}{l}
        0 ,  ~ m < 3 \\
        - \frac{m}{\log(m)}\sum_{k=1}^m p_k \log (p_k), ~ m \ge 3,
   \end{array}\right.
 \label{eq:wadaidx}
\end{equation}
where m is the number of different attractors inside the box. The Wada index is clearly connected to the boundary basin entropy of Eq. \ref{eq:Sb3}, which is the basin entropy computed only on the boxes of the boundary. Indeed, for Wada basins we obtain
\begin{equation}
   S_{bb}=-\dfrac{log N_A}{N_A} W,
 \label{eq:wSbb}
\end{equation}
\textcolor{black}{where $W$ is the average of the Wada index $w$ over the boxes covering the boundary.}
\begin{figure}[!h]
    \centering
    \subfigure[]{\includegraphics[width=0.35\linewidth]{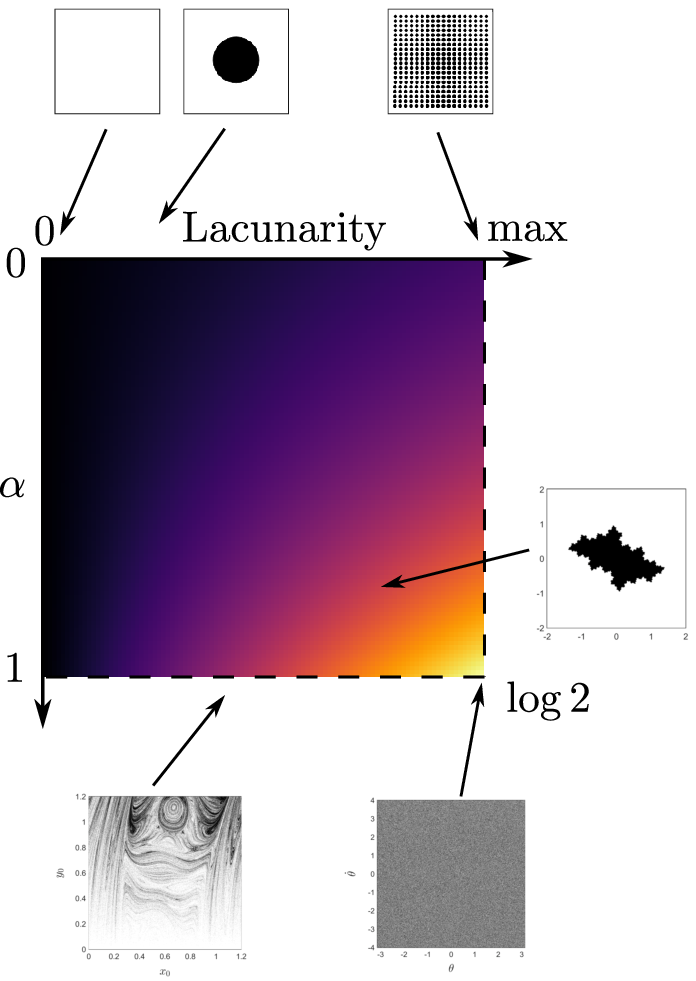}}
    \subfigure[]{\includegraphics[width=0.44\linewidth]{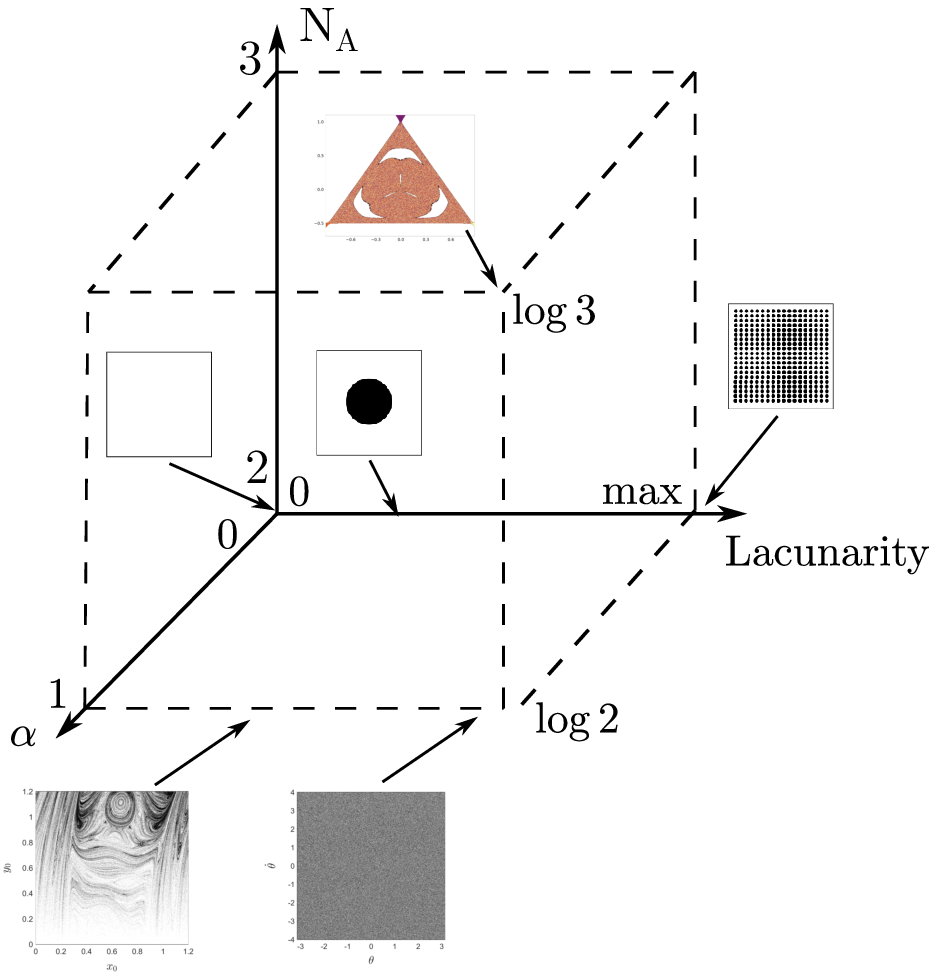}}

    \caption{Interpretation of the basin entropy as a function of two of the important ingredients described in Eq.~\ref{eq:Sb3}. The basin entropy is decomposed artificially into three orthogonal components. Although these components are not independent, this picture is useful to visualize the classification between basins and their associated unpredictability. In (a) we have a heatmap representing the value of the basin entropy on the plane lacunarity-$\alpha$. The coloring is completely arbitrary, but it illustrates the correlation between the basin entropy and these two quantities. \textcolor{black}{The disconnected Fatou set is not included on this plot since its basin entropy varies depending on the resolution chosen for the representation of the points of the set. This may lead to unintuitive situations where basins with higher lacunarity have a lower basin entropy}. In (b) we have extended the plane to a three dimensional cube to include the number of attractors in the analysis. The examples described here, such as the Wada basins obtained from the H\'{e}non-Heiles system, have been placed on their corresponding place.}
    \label{fig:classif}
\end{figure}

This simple relation, which only holds for Wada basins, has been numerically verified for the H\'enon-Heiles Hamiltonian. This paradigmatic 2D open Hamiltonian system is defined by $H=\frac{1}{2}(\dot{x}^2+\dot{y}^2)+\frac{1}{2}(x^2+y^2)+x^2y-\frac{1}{3}y^3$. For an energy above its critical value, it has three exits and the corresponding escape basins show the Wada property~\cite{aguirre_wada_2001}. Varying the total energy of the system, we can find hyperbolic or non-hyperbolic regimes and the uncertainty exponent $\alpha$ changes accordingly, but in all cases the escape basins present the Wada property~\cite{nieto2020measuring}. In Fig.~\ref{fig:NA}(c), we depict the computation of the basin entropy $S_{b}$ (blue line) and the boundary basin entropy $S_{bb}$ (green line) for different values of the energy $E$ in the H\'enon-Heiles system. First, the Eq.~\ref{eq:wSbb} holds numerically, since the dots calculated as $-\dfrac{log N_A}{N_A}W$ perfectly match the boundary basin entropy $S_{bb}$. Second, as the energy approaches its critical value $E_c=1/6$ (vertical dashed line), both the basin entropy $S_b$ and the boundary basin entropy $S_{bb}$ tend to their maximum possible value, in this system $\log 3$ (horizontal dashed line). This result is connected to the total fractalization of the basins taking place at the critical value of the energy~\cite{aguirre2003limit}. At the critical value, the boundary occupies all the phase space, so that all the boxes are in the boundary and therefore $S_b=S_{bb}$. Also, given the $2\pi/3$ rotational symmetry of the system, the three basins are equiprobable. It maximizes the possible uncertainty for $N_A=3$, defined by $S_b=S_{bb}=\log 3$. This example underscores the importance of the Wada property and the Wada index in the unpredictability of a dynamical system.

\section{Discussion}
\label{sec:discussion}
Along the previous lines we have explained the contributions of different ingredients to the basin entropy and, consequently, to the unpredictability associated to basins of attraction in multistable systems. The interesting point is that it opens a way to classify basins of attraction. We can identify some of the most notable types of basins as cases where these ingredients take extreme values. For example, basins with smooth boundaries have $\alpha=1$, which is the maximum possible value for the uncertainty exponent and therefore they are the most predictable in this respect.

On the other side of the spectrum we find riddled basins, which have the minimum possible value of $\alpha=0$. They are the most unpredictable basins in the sense of how the ratio of uncertain initial conditions grows with resolution. Also, riddled basins have large values for the lacunarity, given their Cantor-like disconnected structure.

Concerning the number of attractors, the Wada property maximizes the number of basins separated by one boundary. Nevertheless, the distribution of the different probabilities plays a crucial role too. This can be quantified by the Wada index, and its value often depends on the symmetry of the system. Wada boundaries are fractal boundaries so they always have $\alpha<1$, but there is a case where boundaries separate all possible basins (like Wada basins) and also has $\alpha=0$ (like riddled basins). This kind of basins is called intermingled~\cite{lai1995intermingled,sommerer1996intermingled,alexander1996intermingled} and they maximize the unpredictability in two different ways at the same time. Although the Wada index may vary, intermingled basins are the most unpredictable basins according to their basin entropy.

\textcolor{black}{The basin entropy can be used to detect these extreme cases using the criteria listed in Table \ref{table} for each case. These necessary conditions on $S_b$, $S_{bb}$ and $W$ can help to identify these special basins. However, the basin entropy remains a classification tool since it sets a hierarchy among the basins from the point of view of the unpredictability in the initial conditions space. }

\begin{table}
\begin{center}
\begin{tabular}{ccccc}
\hline
 {Basin type} & \multicolumn{3}{c}{Ingredients values} & {Basin entropy criterion}\\
 \cmidrule{2-4}
 & \parbox[c]{1.5cm}{$\alpha$} & \parbox[c]{1.5cm}{lacunarity} & \parbox[c]{2cm}{$N_A$} & \\
  \hline
 Smooth & 1 & - & - &  $S_{bb} = 0.439$ \cite{puy2021test}\\
 Fractal & $\neq 1$ & - & - &  $S_{bb} \neq 0.439$ \cite{puy2021test}\\
 Wada & $<1$ & - & $\geq 3$ & $S_{bb} = -\dfrac{log N_A}{N_A} W$\\
 Riddled & 0 & max & 2 & $S_b = S_{bb}$ \\
 Intermingled & 0 & max & $\geq 3$ & $S_b = S_{bb} = -\dfrac{log N_A}{N_A} W$  \\
 \hline
\end{tabular}
\caption{\label{table} \textcolor{black}{Summary of the classification of the different basins according to the basin entropy. The main ingredients present in the basin entropy are listed with their typical values. The column on the right expresses a necessary condition on the basin entropy and boundary basin entropy to detect the type of basin involved. This table summarizes the extreme cases,  but the basin entropy can also compare two basins from the perspective of the unpredictability in the phase space.}}
\end{center}
\end{table}
The classification can be visualized graphically in Fig.~\ref{fig:classif}(a) and \ref{fig:classif}(b) where the basin entropy has been decomposed artificially on three independent axis. The examples presented in this work have been represented and located on this graph. Although the lacunarity and the uncertainty exponent are dependent of each other, this figure is illustrative of the main contributions of these ingredients to the basin entropy.

\section*{Conflict of interest}
The authors declare that they have no conflict of interest.

\section*{CRediT authorship contribution statement}
\textbf{Alvar Daza:} Investigation, Visualization, Software, Formal analysis, Writing - original draft. \textbf{Alexandre Wagemakers:} Investigation, Visualization, Software, Formal analysis, Writing - review \& editing. \textbf{Miguel A.F. Sanjuán:} Supervision, Conceptualization, Investigation, Formal analysis, Writing - review \& editing, Funding acquisition.

\section{Acknowledgments}

The authors acknowledge support from the Spanish State Research Agency (AEI) and the European Regional Development Fund (ERDF, EU) under project PID2019-105554GB-I00.

\bibliography{biblio_classification}
\bibliographystyle{model3-num-names}

\end{document}